%% file: IPCCC-TR.tex
\documentclass[12pt] {article}
\pdfoutput=1
\usepackage{epsfig}
\usepackage{amssymb}
\usepackage{amsmath}
\usepackage{multirow}
\usepackage{setspace}
\usepackage{datetime}
\usepackage{amsmath}
\usepackage{algorithm}
\usepackage{algpseudocode}

\usepackage{acronym}
\usepackage{array}
\usepackage{graphicx}
\usepackage{cite}
\usepackage{graphicx}
\usepackage{epstopdf}
\usepackage{caption}
\usepackage{subcaption}
\usepackage{balance}
\usepackage{fancybox}
\usepackage{colortbl}
\usepackage{array}
\usepackage{mystyle}
\usepackage{subcaption}
\usepackage{multirow}
\usepackage[obeyspaces,hyphens,spaces]{url}
\usepackage{threeparttable}

\renewcommand{\baselinestretch}{1}

\providecommand{\keywords}[1]{\textbf{\textit{Index terms---}} #1}

{\begin{list}{}%
         {\setlength{\leftmargin}{#1}}%
         \item[]%
}
{\end{list}}

\newdateformat{monthyeardate}{%
  \monthname[\THEMONTH], \THEYEAR}
\begin{document}
\input{abstractnew}
\newpage
\input{introduction}

\input{background}

\input{Motivation}
\input{proposedsystem}

\input{results}

\input{threats}

\input{relatedwork}

\input{conclusion}

\balance
\bibliography{mabiblio}{}
\bibliographystyle{IEEEtran}
\end{document}

%% file: abstractnew.tex
\title{\huge{ATLAS: An Adaptive Failure-aware Scheduler for Hadoop} }
\author{\begin{tabular}[t]{c@{\extracolsep{10em}}c}
\multicolumn{2}{c}{\hspace*{1em}Mbarka~Soualhia$^{1}$, Foutse~Khomh$^{2}$ and Sofi\`ene~Tahar$^{1}$} \vspace*{2em}\\
$^{1}$Department of Electrical and Computer Engineering,\\
Concordia University, Montr\'eal, Canada  \\
\{soualhia,tahar\}@ece.concordia.ca \vspace*{2em}\\
$^{2}$Department of Software Engineering,\\
\'Ecole Polytechnique, Montr\'eal, Canada \\
foutse.khomh@polymtl.ca \vspace*{3em}\\
\textbf{TECHNICAL REPORT}\\
\date{November 2015}
\end{tabular}}
\maketitle

\newpage
\begin{abstract}
Hadoop has become the \textit{de facto} standard for processing large data in today's cloud environment. The performance of Hadoop in the cloud has a direct impact on many important applications ranging from web analytic, web indexing, image and document processing to high-performance scientific computing. However, because of the scale, complexity and dynamic nature of the cloud, failures are common and these failures often impact the performance of jobs running in Hadoop. Although Hadoop possesses built-in failure detection and recovery mechanisms, 
%
%
several scheduled jobs still fail because of unforeseen events in the cloud environment. A single task failure can cause the failure of the whole job and unpredictable job running times. 
In this paper, we propose ATLAS (AdapTive faiLure-Aware Scheduler), a new scheduler for Hadoop that can adapt its scheduling decisions to events occurring in the cloud environment. Using statistical models, ATLAS predicts task failures and adjusts its scheduling decisions on the fly to reduce task failure occurrences. 
We implement ATLAS in the Hadoop framework of Amazon Elastic MapReduce (EMR) and perform a case study to compare its performance with those of the FIFO, Fair and Capacity schedulers. Results show that ATLAS can reduce the percentage of failed jobs by up to 28\% and the percentage of failed tasks by up to 39\%, and the total execution time of jobs by 10 minutes on average. ATLAS also reduces CPU and memory usages.

\end{abstract}
\keywords{Failure Prediction, Tasks Scheduling, Cloud, Google Clusters, Hadoop, Amazon Elastic MapReduce.}
\newpage
\tableofcontents

%% file: introduction.tex
\section{Introduction}
MapReduce~\cite{MapReduce-Dean2008} has emerged as the leading programming model for large-scale distributed data processing. Hadoop~\cite{AHybrid-Rasooli2012}, the open-source implementation of MapReduce has become the framework of choice on many off-the-shelf clusters in the cloud. It is extensively used in many applications ranging from web analytic, web indexing, image and document processing to high-performance scientific computing and social network analysis. Major large companies like Google, Facebook, Yahoo or Amazon rely daily on Hadoop to perform important data-intensive operations in their data centers. However, because of the scale, complexity and the dynamic nature of cloud environments, failures are common in data centers powering the cloud. Studies \cite{Dean-Experiences2006} show that more than one thousand individual machine failures and thousands of hard-drive failures can occur in a cluster during its first year of service. Several power problems can also happen bringing down between 500 and 1000 machines for up to 6 hours. The recovery time of these failed machines being as high as 2 days. These frequent failures in data centers have a significant impact on the performance of applications running Hadoop~\cite{Dinu-Understanding2012,Kim15}. 
Dinu et al.\cite{Dinu-Understanding2012,Dinu-Hadoop2011} who examined the performance of Hadoop under failures reported that many task failures occur because of a lack of sharing of failure information between the different components of the Hadoop framework. 
\\

\noindent The Hadoop scheduler is a centrepiece of the Hadoop framework. An effective Hadoop scheduler can avoid submitting tasks on fault-prone machines; which would reduce the impact of machine failures on the performance of the applications running Hadoop. However, basic Hadoop scheduling algorithms like the FIFO algorithm, the Fair-sharing algorithm, and the Capacity algorithm only rely on small amount of system information to make their scheduling decisions. They are not equipped with pro-active failure handling mechanisms. Yet, a single task failure can cause the failure of a whole job and unpredictable job running times. 
In our previous work~\cite{Mbarka15} we have shown that it is possible to predict task and jobs scheduling failures in a cloud environment and that such predictions can reduce the percentage of failed jobs by up to 45\%. But, we did not propose an  efficient strategy to reschedule tasks predicted as failed.
\\

\noindent In this paper, we build on that previous work and propose ATLAS (AdapTive faiLure-Aware Scheduler), a new scheduler for Hadoop that adapts its scheduling decisions to events occurring in the cloud environment. Using information about events occurring in the cloud environment (\eg{} resource depletion on a node of the cluster or failure of a scheduled task) and statistical models, ATLAS predicts the potential outcome of new tasks and adjusts its scheduling decisions accordingly to prevent them from failing. 
In addition, ATLAS scheduler introduces novel strategies to reschedule tasks predicted as failed like multiple speculative executions and penalty mechanism.
We implement ATLAS in the Hadoop framework of Amazon Elastic MapReduce (EMR). \emph{To the best of our knowledge, ATLAS is the first scheduler for Hadoop that adapts its scheduling decisions based on predicted failures information}. We perform a case study using multiple single and chained Hadoop
jobs (these jobs are composed of \textit{WordCount}, \textit{TeraGen} and \textit{TeraSort} job units~\cite{Hadoop-Jobs}), to compare the performance of ATLAS with those of the FIFO, Fair and Capacity schedulers. To assess the performance of each scheduler, we compute the total execution times of jobs, the amount of resources used (CPU, memory, HDFS Read/Write), the numbers of finished and failed tasks and the numbers of finished and failed jobs. Using this information, we answer the following research question.\\
\textbf{RQ: Does ATLAS outperforms FIFO, Fair, and Capacity schedulers in terms of execution time, number of finished and failed tasks, number of finished and failed jobs, and resource usage?
}\\
Results show that ATLAS can reduce the percentage of failed jobs by up to 28\%, the percentage of failed tasks by up to 39\%. Although ATLAS requires training a predictive model, we found that the reduction in the number of failures largely compensates for the model training time. In fact, ATLAS even reduces the total execution time of jobs by 10 minutes on average. ATLAS also reduces CPU and memory usages, as well as the number of HDFS Reads and writes.
\\

\noindent \textbf{The remainder of this paper is organized as follows:} Section \ref{sec:background} presents background information about Hadoop. Section \ref{sec:motivation} presents the motivation for this work. Section \ref{sec:proposedsystem} describes our proposed scheduler (\ie{} ATLAS). Section \ref{sec:results} describes our case study and discusses the obtained results. Section \ref{sec:threats} discusses threats to the validity of our work. Section \ref{sec:relatedwork} summarizes the related literature, and Section \ref{sec:conclusion} concludes the paper and outlines some avenues for future works.

%% file: background.tex
\section{Background}
\label{sec:background}
In this section, we briefly introduce MapReduce, Hadoop and its schedulers.
\subsection{MapReduce}
\begin{figure}[th]
\centering
\includegraphics[scale=.45]{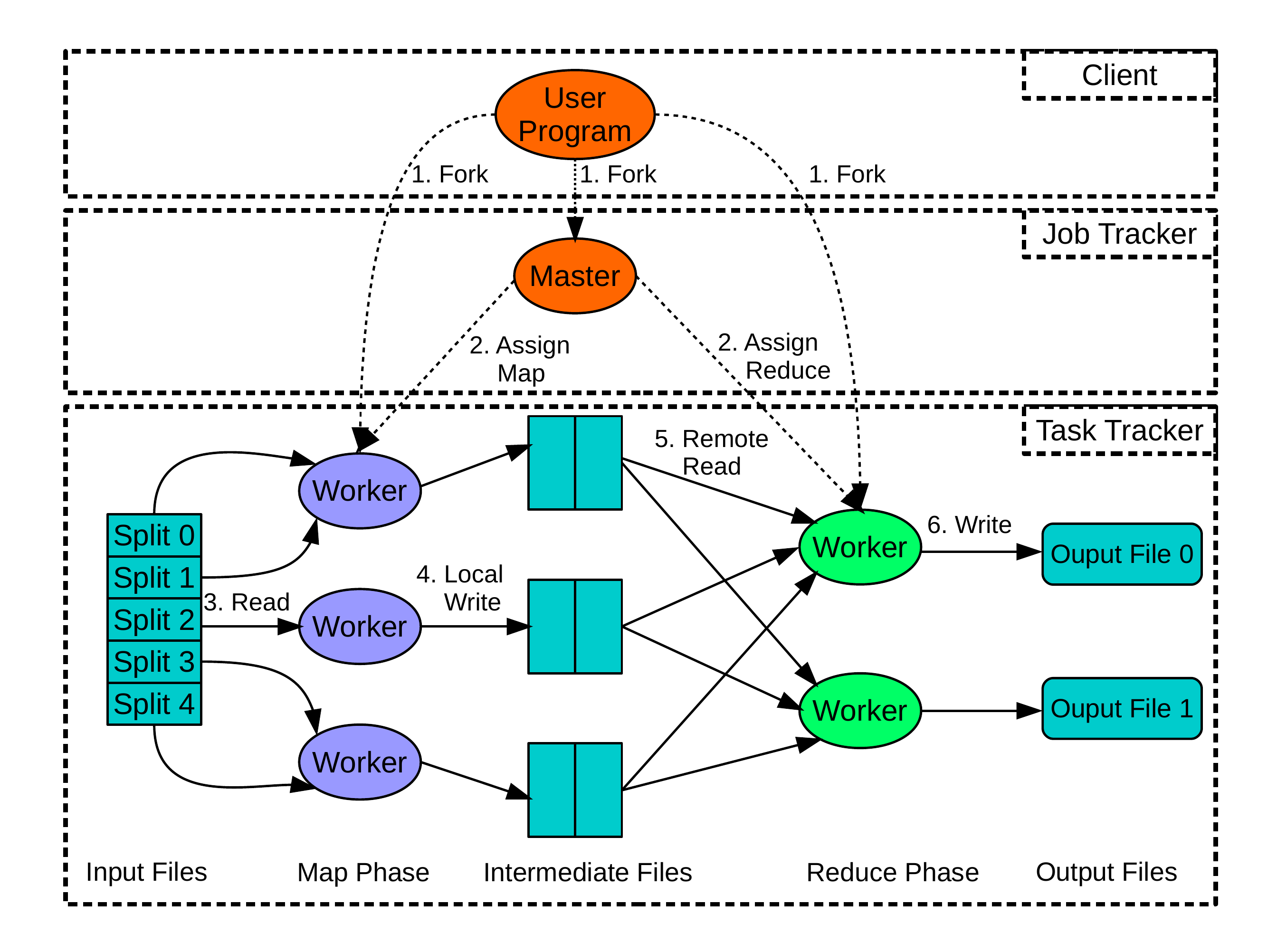}
\caption{An overview of Job Execution in MapReduce \cite{MapReduce-Dean2008}}
\label{Figure:HadoopMapReduce}
\vspace{-5pt}
\end{figure}

MapReduce is a programming model inspired by functional programming languages like Lisp. It is designed to perform parallel processing of large datasets using a large number of computers (nodes)~\cite{Par-Lee2012}. It splits jobs into parallel sub-jobs to be executed on different processing nodes where the data are located instead of sending the data to where the jobs will be executed. A MapReduce job is composed of map and reduce functions and the input data. The map function subdivides the input record into a set of intermediate $<$key, value$>$ pairs. The input data which are called splits represent a set of distributed files that will be assigned to the mappers. The reduce function takes a set of values to process for a same key and generates the output for this key. MapReduce requires a master (known as ``JobTracker") that controls the execution procedure across the mappers (\ie{} worker running a map function) and the reducers (\ie{} worker running a reduce function), using ``TaskTrackers", in order to ensure that all the functions are executed and have their input data as shown in Figure~\ref{Figure:HadoopMapReduce}. 
The intermediate data generated from the map functions are assigned to the reducers based on a key hash function that collects the outputs of mappers having the same key. Each reducer can execute a set of intermediate results belonging to the mappers at a time, then all the generated results will be aggregated after the shuffling step to get the final output once the reducers finish their tasks~\cite{Abstract-Schultz2012}.

\subsection{Apache Hadoop}
Hadoop is a Java-based open source implementation of MapReduce proposed by Doug Cutting and Mike Cafarella in 2005. It has become the de facto standard for processing large data in today's cloud environment~\cite{AHybrid-Rasooli2012}. 
Hadoop is composed of two main units: a storage unit (Hadoop Distributed File System (HDFS)) and a processing unit (MapReduce). It has a Master-Slave architecture: the master node consists of a JobTracker and NameNode. A slave (or worker) node can act as both a DataNode and TaskTracker. Hadoop hides all system-level details related to the processing of parallel jobs (such as the distribution to HDFS file store or error handling), allowing developers to write and enhance their parallel programs while focusing only on the computations issues rather than the parallelism ones.
%
Over the years, a collection of additional frameworks have been proposed to enhance the features of Hadoop. For example, the Apache Pig platform allows creating MapReduce programs for Hadoop using the Pig Latin language and the data warehouse Apache Hive enables data summarization, query, and analysis of large datasets stored in Hadoop's HDFS. 
\subsection{Hadoop Schedulers}
The default scheduling algorithm of Hadoop is based on the First In First Out (FIFO) principle. The JobTracker of Hadoop is responsible for scheduling and provisioning the submitted jobs. These functions were first located in one daemon. Hadoop developers decided to subdivide them into one Resource Manager and and per-application Application Masters to have more flexibility and facilitate the addition of new schedulers. The Resources Manager daemon is called YARN (Yet Another Resources Negotiator) and is responsible for the attribution of resources to applications. 
Facebook and Yahoo! have developed two new schedulers for Hadoop: Fair Scheduler and Capacity scheduler, respectively.
\subsubsection{FIFO Scheduler}
The default scheduler of Hadoop uses a FIFO queue (without any extra configuration). In cloud environment, users' applications are considered as jobs composed of one or more tasks. Each task is a Linux program involving one or multiple processes.
The received jobs are partitioned into sub-tasks which will be loaded into the queue and executed in the order in which they are submitted, regardless of the type and size of the jobs. Although, the FIFO algorithm is easy to implement and grants a full access to resources to the scheduled jobs (which can result in good performances), starvation is possible (especially when there are long run-time jobs waiting in the queue) as a scheduled job can use the entire resource of the cluster for a long time, while others wait in the queue until they time out. 
Moreover, it does not have the ability to control the number of maps and reduces slots for each job and it does not support multi-user executions~\cite{Matchmaking-Cloud2011}.

\subsubsection{Fair Scheduler}
The Fair Scheduler was developed by Facebook to ensure that resources are assigned fairly among different jobs so that all users get on average the required resources available in the cluster over the time. It considers the number of available slots (\ie{} mappers or reducers)
when making decisions about resource allocation. 
It improves the scheduling of the jobs by assigning the minimum number of slots while guaranteeing a high level of service to support Quality of Service (QoS) requirements. By default, its scheduling decisions are based on memory distribution fairness. However, it can support both CPU and memory allocation according to the notions of Dominant Resource Fairness proposed by Ghodsi \emph{et al.}~\cite{Dominant-Ghodsi2011}. It supports multi-user execution in one cluster unlike the default FIFO scheduler of Hadoop. Moreover, it can use the priorities assigned to users applications as a factor to determine the required resources. Therefore, it can guarantee that long jobs will not be starving, by optimizing their waiting time in the queue~\cite{Improving-Yuting2013}.

\subsubsection{Capacity Scheduler}
The Capacity scheduler was originally implemented by Yahoo!. It supports multi-user execution within one cluster and allows large number of users to execute their jobs fairly over time. In fact, it divides the cluster into multiple queues with configurable capacity (\ie{} CPU, memory, disk, etc.). It assigns the minimum number of available slots to the received jobs in the queues using FIFO scheduling principles. The queues support jobs priorities and guarantee that there is a limit on the allocated resources to all users in order to prevent some jobs from using all available resources in the queue where they are assigned. In addition, it improves resources sharing among the queues in order to fairly distribute the available cluster capacity among users rather than among the submitted jobs (Fair Scheduler). The capacity scheduler relies only on memory capacity allocation to optimize the execution time and the throughput of the data-intensive jobs within Hadoop large clusters~\cite{Enhancement-Raj2012}.

%% file: Motivation.tex
\section{Motivation}
\label{sec:motivation}
\subsection{Limitation of Current Hadoop's Implementation}
Dinu \emph{et al.} analysed the behavior of the Hadoop framework under different types of failure and found that TaskTracker and DataNode failures are very important since they affect the availability of jobs input and output data~\cite{Dinu-Understanding2012,Dinu-Hadoop2011}. In addition, these failures can cause important delays during the execution of HDFS read and write procedures. 
Their experiments showed that a single failure can lead to unpredictable execution time; for example the average execution time of a job, which is 220s, can reach 1000s under a TaskTracker failure and 700s under a DataNode failure. Moreover, they claimed that the recovery time of the failed components (such as TaskTracker or DataNode) in Hadoop can be long and can cause jobs delays which may affect the overall performance of a cluster. As an illustration, let's consider a TaskTracker that sends heartbeats to the JobTracker every 10 minutes (this is the default value), if a failure occurs in the TaskTarcker within the first minute after a communication with the JobTracker, all the tasks assigned to this TaskTracker will fail and the JobTracker will notice these failures only after about 9 minutes, resulting in a delay in the rescheduling of the failed tasks and an increase of the execution time of the job. Even if tasks are speculatively executed to prevent their full rescheduling in the event of a failure, there is still a cost associated with this replication (the resource spent on the speculative executions)~\cite{KoMaking2010}.\\ 

\noindent In addition, DataNode failures have a large impact on the start up time of speculative task executions. 
This is because of the statistical nature of the speculative execution algorithm which is based on data about task progress. In fact, if a task was making good progress and suddenly fails because of a DataNode failure, its speculative execution will start with a delay (\ie{} later than speculative executions of straggler tasks), since Hadoop expected a normal behavior from that task. 
Also, many map and reduce tasks may fail because they exceeded the number of failed attempts allowed by the TaskTracker. For simplicity and scalability, each computing node in Hadoop manages failure detection and recovery on its own and hence the launched tasks can not share failure information between them. Therefore, multiple tasks, including the speculative tasks, may fail because of an error already encountered by a previous task (it can even be a task belonging to the same job)~\cite{Dinu-Understanding2012}. 

\subsection{Problem Formulation}
\begin{figure}[th]
\centering
\includegraphics[ scale=.6]{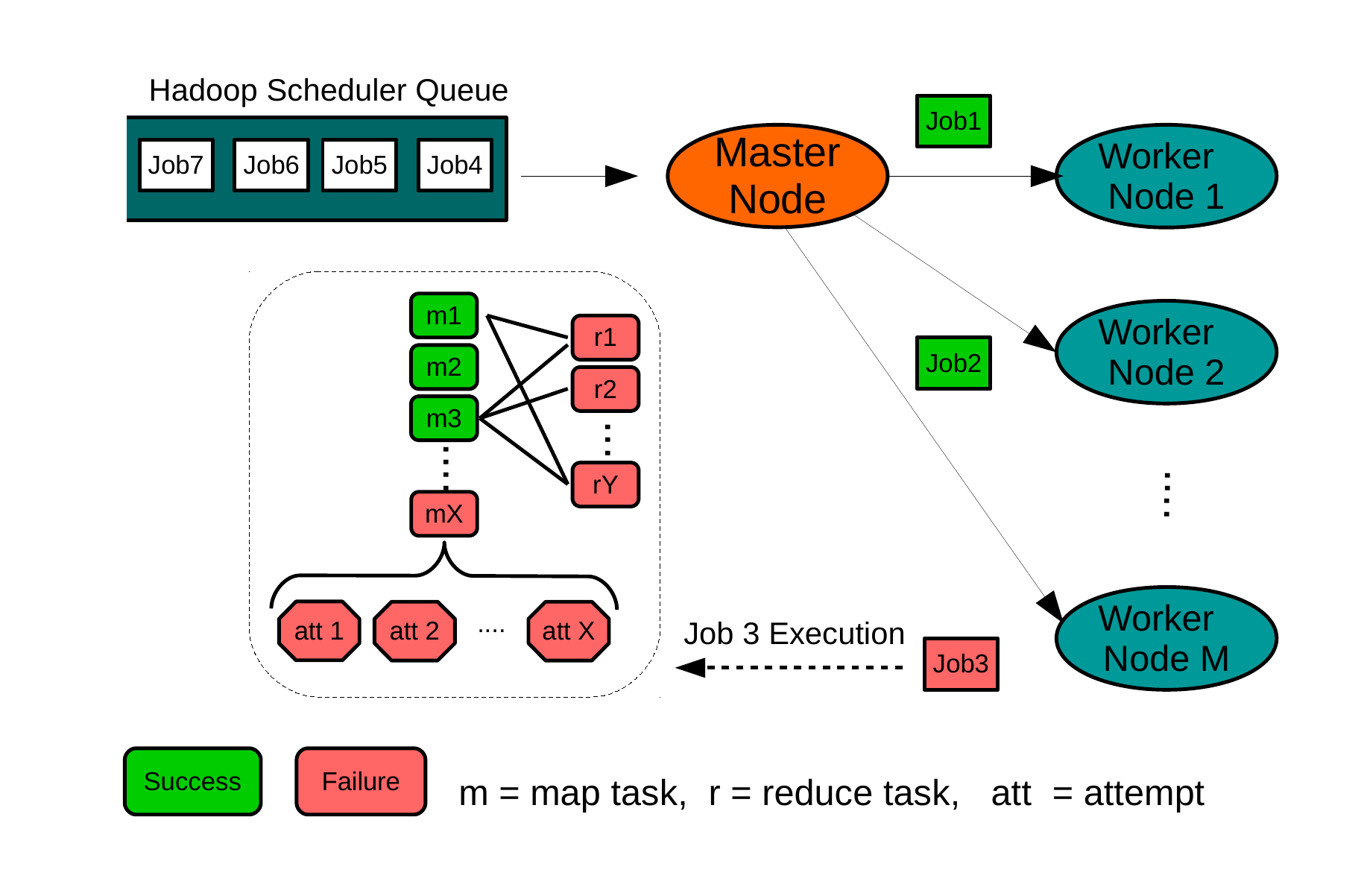}
\vspace{-10pt}
\caption{Example of Hadoop Job Failure}
\label{Figure:exampleofjobfailure}
\end{figure}

Let's consider \textit{N} jobs submitted to Hadoop, where each job is composed of \textit{X} map tasks and \textit{Y} reduce tasks. Let's assume that each job is using \textit{R(CPU, Memory, HDFS Read/Write)} resources from \textit{M} machines in an Hadoop cluster. Each map/reduce task is allowed a maximum number of scheduling attempts: each new task is assigned to a node and if it fails it can be rescheduled multiple times either on the same node or on another available node. When a task exceeds its maximum number of scheduling attempts allowed by the TaskTracker, the task is considered to be failed, otherwise it is finished successfully.
Because of the dependency between map and reduce tasks, if either a map or a reduce task fails, the whole job to which the task belongs will fail as well, even if all the other tasks in the job were completed successfully. For example, in Figure~\ref{Figure:exampleofjobfailure}, \textit{Job3} failed because one of its map tasks failed (because it exceeded its maximum number of scheduling attempts). As a consequence of the failure of this map task, all reduce tasks were failed automatically.
\vspace{-10pt}

\noindent More formally, if $S(job)$ is the outcome of an executed job; $S(MapAtt_{ip})$ the status of a $map_i$ after the $p^{th}$ attempt (we give a value of 1 when an attempt is successful and 0 otherwise) and $S(ReduceAtt_{jq})$ the status of $reduce_j$ after the $q^{th}$ attempt. If \textit{K} and \textit{L} are the maximum numbers of scheduling attempts allowed for map and reduce tasks respectively:
\begin{equation}
\small{S(job)= [\prod_{i=1}^X(\sum_{p=1}^KS(MapAtt_{ip}))]* [\prod_{j=1}^Y(\sum_{q=1}^LS(ReduceAtt_{jq}))]}
\end{equation}

\noindent Given that the execution time of a task is the sum of execution times of all its launched attempts (both the finished and the failed attempts), the more a task experience failed attempts, the longer its execution time will be. This delay in the execution of the tasks will also translate into longer execution times for the job (to which the tasks belong) and larger resources usages. 
More specifically, if \textit{T(job)} is the total execution time of a job composed of a set of $A=\{map_i\}_{i\in X}$ map tasks and a set of $B=\{reduce_j\}_{j\in Y}$ reduce tasks. If $T(MapAtt_{ip})$ and $T(ReduceAtt_{jq})$ are respectively the execution times of the $map_i$ and $reduce_j$ tasks during the attempts $p$, $q$ respectively:
\begin{equation}
\small{T(job)=Max_{A}(\sum_{p=1}^KT(MapAtt_{ip})) + Max_{B}(\sum_{q=1}^LT(ReduceAtt_{jq})) }
\end{equation}


\noindent Therefore, we believe that it is very important to reduce the number of tasks failed attempts if we want to improve the overall performance of an Hadoop cluster. By reducing the number of tasks failed attempts, one will reduce the turnaround time of jobs running in the cluster. 
If one can predict task failure occurrences and adjust scheduling decisions accordingly to prevent failures from occurring, one may be able to reduce the number of tasks failed attempts. In our previous work~\cite{Mbarka15} we have shown that it is possible to achieve such predictions. In the following Section~\ref{sec:proposedsystem}, we build on our previous work and propose ATLAS, a scheduling algorithm that adapts its scheduling decisions based on predicted failures information.

%% file: proposedsystem.tex
\section{ATLAS: an Adaptive Failure-aware Scheduler for Hadoop}
\label{sec:proposedsystem}
\subsection{Proposed Methodology}
\begin{figure}[th] 
\centering
\includegraphics[scale=.35]{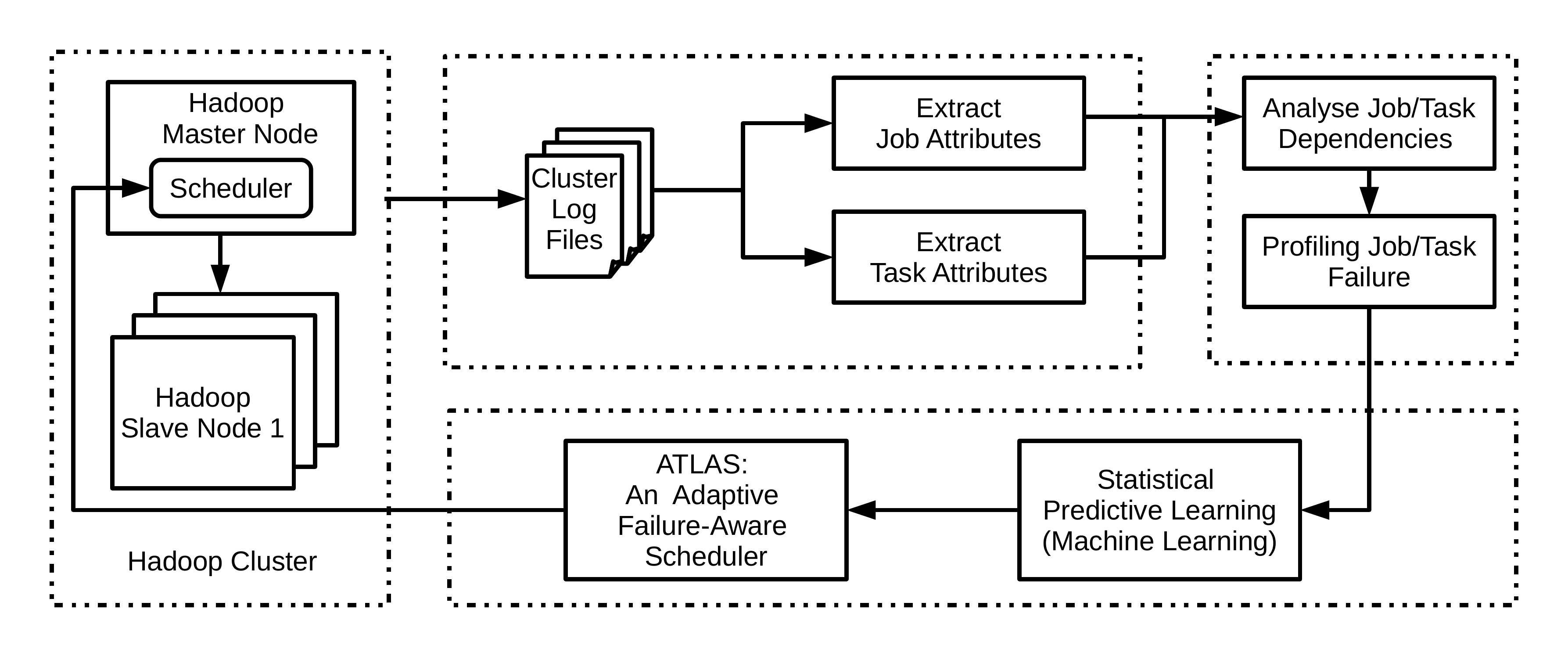}
\vspace{-15pt}
\caption{Overview of Our Proposed Methodology}
\label{Figure:overview}
\vspace{-5pt}
\end{figure}

In this section, we present our scheduler ATLAS, which can reduce the number of tasks failed attempts by predicting task scheduling outcomes and adjusting scheduling decisions to prevent failure occurrences. 
We describe the approach followed to analyse Hadoop's log files and build task failure predictive models. Figure~\ref{Figure:overview} presents an overview of this approach. First, we run different jobs on Hadoop cluster in order to get trace of data about previously executed tasks and jobs. Next, we analyse log files obtained from Amazon EMR Hadoop Clusters and extract jobs and tasks main attributes. Next, we analyse correlations between tasks attributes and tasks scheduling outcomes. Finally we apply statistical predictive learning techniques to build task failures prediction models. 
The remainder of this section elaborates more on each of these steps.
\subsubsection{Extraction of Tasks/Jobs Attributes}
\label{dataextraction}
First, we run different workload in parallel including single and chained jobs on Amazon EMR Hadoop clusters. We run different single jobs in parallel such as \textit{WordCount}, \textit{TeraGen}, \textit{TeraSort}~\cite{Hadoop-Jobs} to get different workload on several machines. In addition, we run chained jobs (sequential, parallel and mix chains) composed of \textit{WordCount}, \textit{TeraGen}, \textit{TeraSort} jobs to get different types of job running on the cluster. Also, we vary the size of the used jobs (number of map and reduce tasks, number of jobs in a chained job). These jobs represent different job pattern similar to the ones running in real world applications. We implemented a bash script to extract job/task attributes as described in Table~\ref{Table:Jobs:Tasks:Description}. For each job we extracted: \textit{job ID, priority, execution time, number of map/reduce, number of local map/reduce tasks, number of finished/failed map/reduce tasks and the final status of the job}. \\For each task we extracted the following information: \textit{job ID, task ID, priority, type, execution time, locality, execution type, number of previous finished/failed attempts of the task, number of reschedule events, number of previous finished/failed tasks, number of running/finished/failed tasks running on the TaskTracker, the amount of used resources (CPU, Memory and HDFS Read/Write) and the final status of the task}.
\begin{table}[!h]
\caption{Jobs and Tasks Attributes}
\vspace{-5pt}
\label{Table:Jobs:Tasks:Description}
\centering \scriptsize
\begin{tabular} {|p{3cm} |p{6cm}|p{5cm}| } 
\hline
\textbf{Job/Task Attributes} & \textbf{Description} & \textbf{Rationale}\\  
\hline \hline
Job/Task ID & Immutable and unique identifier for a job/task & Used to identify a job/task\\ \hline
Type & Type of a task & It represents the type of a task: Map or Reduce\\ \hline
Priority & Preemption type of a task/job & Used to capture task/job priority to access resources \\ \hline
Locality/Execution Type & Locality/Execution type of a task & Used to capture the fact that a task was launched locally/speculatively or not\\ \hline
Execution Time  & Time between submission date and date when task is finished/failed& Used to capture the execution time of a task\\ \hline
Nbr Finished and Failed Tasks& Number of finished and failed tasks & Used to capture the proportion of finished/failed tasks\\ \hline
Nbr Pre. Finished and Failed Attempts& Number of previous finished and failed attempts & Used to capture failure events dependent on a task\\ \hline
Nbr of Reschedule Events& Number of reschedule events of a failed task & Used to capture the number of times that a failed task was rescheduled \\ \hline
Nbr of Finished, Failed and Running Task TT& Number of finished, failed and running tasks on TaskTracker & Used to capture failure events on the same TaskTracker \\ \hline
Available resources on  TT& Amount of available resources on TaskTracker & Used to capture the availability of resources on TT \\ \hline
Total Nbr of Tasks of a Job & Total number of tasks within a job & Used to capture the distribution of tasks within the jobs \\ \hline
Used CPU/RAM/ HDFS R/W & Used CPU, RAM and Disk Space for a task & Used to capture the usage of resources \\ \hline
Final Status  &  Final state on a scheduling life-cycle & Used to describe the processing outcome of a task/job\\
\hline
\end{tabular}
\vspace{-7pt}
\end{table}
\subsubsection{Profiling of Tasks/Jobs Failure}
To identify the correlation between job/task scheduling outcome and their attributes, we analyse the dependencies between the jobs and tasks and perform a mapping between the failed ones and their attributes. Second, we checked the obtained data to identify the most relevant attributes that impact the final scheduling outcome of task/job by removing the ones having unchanged value or null value. This step is preliminary to the following one.
\subsubsection{Statistical Predictive Learning}
\label{models}
We aim to explore the possibility to predict a potential task failure in advance based on its collected attributes and machine learning techniques. We believe that if we can share the failure information between tasks in advance, we can prevent the occurrence of the predicted failure and reschedule them on appropriate clusters to ensure their timely and successful completion.
To do so, we choose several regression and classification algorithms in \textit{R}~\cite{[23]} to build models: GLM (General Linear Model), Random Forest, Neural Network, Boost, Tree and CTree (Conditional Tree). GLM is an extension of linear multiple regression for a single dependent variable. It is extensively used in regression analysis. Boost creates a succession of models iteratively, each model being trained on a data set in which points misclassified by the previous model are given more weight. All the successive models are weighted according to their success and their outputs are combined using voting or averaging to create a final model. Neural networks are graphs of interconnected nodes organised in layers. The predictors (or inputs) form the bottom layer, and the forecasts (or outputs) form the top layer. Decision Tree is a widely used classification approach for predicting binary outcomes. CTree is a different implementation of Decision Tree. Based on Decision Tree, Leo Breiman and Adele Culter developed Random Forest, which uses a majority voting of decision trees to generate classification (predicting, often binary, class label) or regression (predicting numerical values) results~\cite{[23]}. Random Forest offers good out-of-the-box performance and has performed very well in different prediction benchmarks~\cite{[23]}. The algorithm yields an ensemble that can achieve both low bias and low variance~\cite{[23]}.\\

\noindent We use different training and testing data set for both jobs and tasks. We collected data related to 70,000 jobs and 180,000 tasks from the Hadoop cluster we used in our experiments as described in Section~\ref{dataextraction}. The log data were collected over a fixed period of time of 10 minutes. The training time is related to the steps of training process and not to the complexity of the running jobs. 
We apply 10-fold random cross validation to measure the accuracy, the precision, the recall and the error of the prediction models~\cite{[23]}. The accuracy is $\frac{TP + TN}{TP+TN+FP+FN}$, the precision is $\frac{TP}{TP+FP}$, the recall is $\frac{TP}{TP+FN}$, and the error is $\frac{FP + FN}{TP+TN+FP+FN}$,   where $TP$ is the number of true positives, $TN$ is the number of true negatives, $FP$ is the number of false positives, and $FN$ is the number of false negatives.
In the cross validation, each data set is randomly split into ten folds. Nine folds are used as the training set, and the remaining fold is used as the testing set.

\subsection{The ATLAS Scheduling Algorithm}
\vspace{-4pt}
ATLAS aims to provide better scheduling decisions for predicted failed tasks, in order to ensure their successful execution. A scheduling decision may require either assigning the tasks to other TaskTrackers with enough resources or waiting for some other tasks to be finished. We designed ATLAS using the predictive model that provided the best results in terms of precision and accuracy when predicting tasks scheduling outcomes. ATLAS integrates with any Hadoop's base scheduler (like FIFO, Fair, Capacity, etc). In fact, when tasks are predicted to succeed, ATLAS relies on Hadoop's base scheduler to make its scheduling decision.
Algorithm~\ref{ProposedAlgo} presents ATLAS in details. The main algorithm is composed of 3 main parts: (1) a task failure prediction algorithm, (2) an algorithm to check the availability of resources, and (3) a task rescheduling algorithm (for potential failed tasks).\\

\noindent We implemented a procedure to collect the attributes of the tasks (map/ reduce) described in Table~\ref{Table:Jobs:Tasks:Description}. The attributes of the tasks represent the predictors of our models.
Using the values of these attributes, the failure prediction algorithm predicts whether a task will be finished or failed if executed. The response of the trained models is a binary variable taking \textit{``True"} if a scheduled task succeeds and \textit{``False"} if it fails. We implemented two different prediction algorithm for the mappers and the reducers since they have different input parameters.
Next, if a task is predicted to succeed, our algorithm checks the availability of the TaskTracker and DataNode to verify if they are activated or not, since we noticed that the scheduler may assign tasks to a dead TaskTracker, because of the predetermined frequency of heartbeats between the JobTracker and TaskTracker (between two heartbeats a JobTracker has no means to know that a TaskTracker is dead).\\
Moreover, we implemented a procedure to modify the time spent between two successive heartbeats based on collected information about TaskTracker failures. This procedure is running in parallel along with ATLAS. So, if there are very frequent TaskTracker failures (\ie{} if more than 1/3 of TaskTrackers were failed between two heartbeats), the time between two heartbeats will be decreased in order to detect faster node failures and reschedule tasks early on, on other alive nodes. 
This time is decreased each time by half of the previous time elapsed between two heartbeats (\ie{} the value was 10 min, then it will be decreased to 5 min) until reaching a minimum value. The minimum value in our experiment is 2 min.   
If there are less TaskTracker failures (\ie{} less than 1/3 of the workers), this time will be increased in order to reduce the cost associated with communication between the JobTracker and TaskTracker. 
The value of heartbeat is adjusted on the fly according to events related to TaskTracker failures.\\

\noindent After checking the TaskTracker, ATLAS checks if there are enough slots on the selected TaskTracker or not since some tasks may fail because of a high number of concurrent tasks on a TaskTracker.
If an assigned task is predicted to fail but there are enough available resources in the cluster, ATLAS will launch the task speculatively on many nodes (specifically the ones that are not very distant) that have enough resources, in order to speed up the execution of the task and increase the chances of success of the task.
All the decisions made by the ATLAS scheduler are controlled by a time-out metric from Hadoop's base scheduler. Hence, if a task reaches its time-out, its associated attempt will be considered as failed and the task will be rescheduled again but with a low priority. We rely on a penalty mechanism to manage the priority of the tasks. We assign a penalty to tasks causing delays to other tasks and tasks that are predicted to fail multiple times. This penalty reduces their execution priority, causing them to wait in the queue until enough resources are available to enable their speculative execution on multiple nodes. 
In Algorithm~\ref{ProposedAlgo}, we denote \textit{JobTracker} as \textit{JT}, \textit{TaskTracker} as \textit{TT} and \textit{DataNode} as \textit{DN}.
\begin{algorithm}
\caption{: The ATLAS Scheduling Algorithm}
\label{ProposedAlgo}
\begin{algorithmic}[1]
\scriptsize
{
\If { \textit{(TypeofTask(Task) == "Map")}}
\State \textit{/* Collect attributes of Task listed in Table~\ref{Table:Jobs:Tasks:Description} */}
\State \textit{Attributes = Collect-Attributes-Map(Task)}
\State \textit{/* Learning Algorithm will predict if Task will be finished/failed */}
\State \textit{Predicted-Status = Predict-Map(Task, Attributes)}
\Else
\State \textit{Attributes = Collect-Attributes-Reduce(Task) }
\State \textit{Predicted-Status = Predict-Reduce(Task, Attributes)}
\EndIf

\If {\textit{(Predicted-Status == "SUCCESS")}}
\State \textit{/* Check whether the TT and DN are dead or alive */}
\State \textit{Check-Availability(TT,DN)}
       \If {\textit{(TT and DN are available)}}
        \State \textit{/* Test if the TT have enough slots to serve Task */}
        \State \textit{Check-Availability-Slots(Task,TT)}
               \If {\textit{(Slots are available in TT)}}
                \State \textit{Execute(Task,TT)}
                \Else
                \State \textit{Wait Until Free Slots in TT and Time-Out Not Reached}
                \If {\textit{(Time-Out is Reached )}}
                   \State \textit{/* Resubmit Task since it will fail in such conditions */}
                   \State \textit{Send to Queue + Penalty }
                \Else
                \State \textit{/* Execute Task in the TaskTracker TT */}
                \State \textit{Execute(Task,TT)}
               \EndIf
               \EndIf
        \Else

        \While {\textit{(TT/DN not activated and Time-Out Not Reached )}}
        \State \textit{/* Send a HeartBeat to the JT to activate TT/DN */}
        \State \textit{Notify JT to Activate TT/DN}
        \EndWhile
         \If {\textit{(Time-Out is reached)}}
           \State \textit{Send to Queue + Penalty }
          \Else
           \State \textit{Execute(Task,TT)}
         \EndIf

       \EndIf

\Else
\If {\textit{(There are Enough Resources on Nodes)}}
\State \textit{/* Launch Many Speculative Instance of Task to increase the probability of its success/}
\State \textit{Execute-Speculatively(Task,N)}
\EndIf

\EndIf

}
\end{algorithmic}
\end{algorithm}

%% file: results.tex
\section{Evaluation}
\label{sec:results}
In this section we presents the design of our case study aimed at assessing the effectiveness of the ATLAS scheduler. 
\subsection{Setup of the Case Study}
We instantiated 15 Hadoop machines in Amazon EMR. We set one machine to the role of master, another one to the role of secondary master (to replace the master in case of failure) and 13 machines to the role of slaves.
We selected 3 different types of machines to have heterogeneous environment which represents a real world environment hosting real and different machines.
The three types of Amazon EMR machines are \textit{m3.large}, \textit{m4.xlarge} and \textit{c4.xlarge}~\cite{AmazonInstances}. 
Details about their characteristics are listed in Table~\ref{Table:CharacteristiqueMachines}.
We choose these types of machine since they can support different workloads and they were widely used in the literature to test many systems in cloud environment. In addition, different types of Amazon EMR instance allow to have a real world cluster
where different types of machines are used.\\

\noindent We used the AnarchyApe tool described in~\cite{Faghri-Failure2012} to create different failure scenarios in Hadoop nodes such as TaskTarcker and DataNode failures, slowdown or drop in the network, loss of some job data, failure of tasks/jobs, and so on.
Specifically, we killed/suspended DataNodes/TaskTrackers; disconnected/slowed/dropped network; and randomly killed/suspended threads within DataNodes/TaskTrackers in the running executions.
To specify the amount of failures to be injected in the Hadoop clusters, we performed a quantitative analysis of failures in the public Google Traces~\cite{Mbarka15}, which contains different types of jobs (including Hadoop jobs). The public Google Traces provides data about task and job failures in real world Google clusters. 
We found that more than 40\% of the tasks and jobs can be failed~\cite{Mbarka15}. Therefore, in our case study,
we performed different simulations of varying the injected failure rates, with a maximum failure rate of 40\%.\\

\noindent To generate the jobs, we used the \textit{WordCount}, \textit{TeraGen} and \textit{TeraSort} examples provided by Apache, as job unit to create single or chained jobs as described in Section~\ref{dataextraction}.
The generated jobs represent different job pattern from real world applications.
These jobs have different input files (\eg{} books\footnote{http://www.gutenberg.org/wiki/Main\_Page})
that we downloaded to have large set of data to process. For each single job, we decided on the number of map and reduce tasks using information about the number of HDFS blocks in the input files. For example, we had jobs with 10 map tasks and 15 reduce tasks.
For each chained job, we decided on the number of job unit within each chain (\eg{} 3 jobs, 20 jobs), the type of used job (\eg{} \textit{WordCount}, \textit{TeraSort}) and the structure of the jobs (\eg{} sequential, parallel and mix chains).
For each type of Hadoop scheduler (\ie{} FIFO, Fair, Capacity), we generated several single and chained jobs and collected their execution logs to build the task failure prediction model required by ATLAS. 
We built prediction models using all the classification algorithms described in Section~\ref{models} and assessed their performances following the 10-fold cross-validation approach described in Section~\ref{models}.
Also, we retrained the prediction models in the instantiated Amazon EMR machines which represent a cloud environment where drastic changes may occur because of unreliable machines. This step was performed each 10 min to make the proposed system more robust.
We implemented ATLAS using the prediction model that achieved the best performance, and compared the performance of Hadoop equipped respectively with the FIFO scheduler, the Fair scheduler, the capacity scheduler, and our proposed ATLAS scheduler integrated with these 3 existing schedulers. All the comparisons were done using the exact same jobs and data. We measured the performance each Hadoop's scheduler using the total execution times of jobs, the amount of resources used (CPU, memory, HDFS Read/Write), the numbers of finished and failed tasks, and the numbers of finished and failed jobs.

\begin{table}
\caption{Amazon EC2 Instance Specifications}
\vspace{-5pt}
\centering \scriptsize
\label{Table:CharacteristiqueMachines}
\begin{tabular}{|c|c|c|c|c|}
\hline
{\bf \begin{tabular}[c]{@{}c@{}}Machine \\ Type\end{tabular}} & {\bf vCPU\textsuperscript{*}} & {\bf \begin{tabular}[c]{@{}c@{}}Memory\\ (GiB)\end{tabular}} & {\bf \begin{tabular}[c]{@{}c@{}}Storage \\ (GB)\end{tabular}} & {\bf \begin{tabular}[c]{@{}c@{}}Network \\ Performance\end{tabular}} \\ \hline
{\bf m3.large}                                                & 1          & 3.75                                                         & 4                                                             & Moderate                                                             \\ \hline
{\bf m4.xlarge}                                               & 2          & 8                                                            & EBS-Only\textsuperscript{+}                                                      & High                                                                 \\ \hline
{\bf c4.xlarge}                                               & 4          & 7.5                                                          & EBS-Only\textsuperscript{+}                                                      & High                                                                 \\ \hline
\end{tabular}
\begin{tablenotes}
      \footnotesize
      \item   \textsuperscript{*} Each vCPU is a hyperthread of an Intel Xeon core~\cite{AmazonInstances}.
      \item   \textsuperscript{+} Amazon EBS is a block-level storage volume attached to a running Amazon EC2 instance~\cite{AmazonInstances}.
   \end{tablenotes}
   \vspace{-15pt}
\end{table}

\subsection{Case Study Results}
\subsubsection{Prediction Algorithms}\label{predictionresults}
\noindent Table~\ref{Table:crossvalisationresults} summarises the performance of the six prediction models applied on data collected from the schedulers' logs. This result shows that the Random Forest algorithm achieves the best precision, recall, accuracy and error when predicting the scheduling outcome of the Map and Reduce tasks for the three studied schedulers (FIFO, Fair and Capacity).
This is because Random-Forest algorithm uses the majority voting on decision trees to generate results which makes it robust to noise, resulting usually in highly accurate predictions.
For map tasks, a Random Forest model can achieve an accuracy up to 83.7\%, a precision up to 86.4\%, a recall up to 94.3\% and an error up to 21.7\%.
The total execution time of the 10-fold cross-validation was 25.97 ms.
For reduce tasks, the Random Forest model achieved an accuracy up to 95.3\%, a precision up to 98.1\%, a recall up to 95.9\% and an error up to 16.6\%.
The total execution time of the evaluation of Random Forest for reduce tasks was 35.65 ms.
Also, we noticed that Random Forest is achieving these results in a acceptable time compared to the other algorithms (for example, the Boost algorithm can take up to 297.29 ms which can affect the performance of the scheduler).
In addition, we also found a strong correlation between the number of running/finished/failed tasks on a TaskTracker, the locality of the tasks, the number of previous finished/failed attempts of a task, and the scheduling outcome of the task. More specifically, tasks characterized by multiple past failed attempts, many concurrent tasks (running on the same TaskTracker) that experienced multiple failed attempts, have a high probability to fail in the future.
\begin{table*}[ht!]
\caption{Accuracy, Precision, Recall, Error (\%) and Time(ms) for different Algorithms: (Random 10-fold Cross-validation)}
\vspace{-5pt}
\centering \tiny
\label{Table:crossvalisationresults}
\begin{tabular}{|> {\centering\arraybackslash}p{0.8cm}|> {\centering\arraybackslash}p{1.2cm}|> {\centering\arraybackslash}p{0.3cm}|> {\centering\arraybackslash}p{0.3cm}|> {\centering\arraybackslash}p{0.3cm}|> {\centering\arraybackslash}p{0.3cm}|> {\centering\arraybackslash}p{0.6cm}|> {\centering\arraybackslash}p{0.3cm}|> {\centering\arraybackslash}p{0.3cm}|> {\centering\arraybackslash}p{0.3cm}|> {\centering\arraybackslash}p{0.3cm}|> {\centering\arraybackslash}p{0.6cm}|> {\centering\arraybackslash}p{0.3cm}|> {\centering\arraybackslash}p{0.3cm}|> {\centering\arraybackslash}p{0.3cm}|> {\centering\arraybackslash}p{0.3cm}|> {\centering\arraybackslash}p{0.6cm}|}
\hline
{\bf Task}                         & {\bf Scheduler}      & \multicolumn{5}{c|}{{\bf FIFO}}                                     & \multicolumn{5}{c|}{{\bf Fair}}                                     & \multicolumn{5}{c|}{{\bf Capacity}}                                 \\ \hline \hline
\multirow{7}{*}{{\bf Map}}    & {\bf Algo.}      & {\bf Acc.}     & {\bf Pre.}     & {\bf Rec.}     & {\bf Err.} & {\bf Time}      & {\bf Acc.}     & {\bf Pre.}     & {\bf Rec.}     & {\bf Err.}   & {\bf Time}         & {\bf Acc.}     & {\bf Pre.}     & {\bf Rec.}     & {\bf Err.} & {\bf Time}           \\ \cline{2-17}
                                   &
{\bf Tree}           & 68.6           & 85.3           & 74.3       &  31.4  & 12.34      &       92.6           & 85.2           & 62.1        &  7.4 & 9.14    &         68.7           & 85.7           & 73.4         & 31.3 & 58.18            \\ \cline{2-17}
                                   &
{\bf Boost}          & 75.9           & 86.7           & 78.5         & 24.1 & 180.51           & 65.9           & 85.6           & 73.5        &  34.1 & 199.80        &   72.5           & 85.3           & 71.3        &  27.5 & 280.56          \\ \cline{2-17}
                                   &
{\bf Glm}            & 63.5           & 87.2           & 72.4      &  36.5   & 9.43            & 67.4           & 88.6           & 65.8           & 32.6 & 13.34    &        62.1           & 83.6           & 61.8         &  37.9 & 16.01           \\ \cline{2-17}
                                   &
{\bf CTree}          & 65.9           & 87.5           & 65.2        & 34.1  & 15.61           & 62.6           & 85.2           & 69.3           & 37.4 & 16.38           & 61.8           & 79.5          &   61.2    &   38.2    & 17.53           \\ \cline{2-17}
                                   &
{\cellcolor[gray]{.8} \bf R.F.}  & {\cellcolor[gray]{.8} \bf  83.7} & {\cellcolor[gray]{.8} \bf  86.4} & {\cellcolor[gray]{.8} \bf  94.3} & {\cellcolor[gray]{.8} \bf 16.3 } & {\cellcolor[gray]{.8} \bf  23.53}  & {\cellcolor[gray]{.8} \bf  79.8} & {\cellcolor[gray]{.8} \bf  83.9} & {\cellcolor[gray]{.8} \bf 94.0} & {\cellcolor[gray]{.8} \bf 20.2 } & {\cellcolor[gray]{.8} \bf  25.91}  & {\cellcolor[gray]{.8} \bf  78.3} & {\cellcolor[gray]{.8} \bf  85.2} & {\cellcolor[gray]{.8} \bf  89.2} & {\cellcolor[gray]{.8} \bf 21.7 } & {\cellcolor[gray]{.8} \bf  25.97} \\ \cline{2-17}
                                   &
{\bf N.N.} & 65.8           & 85.8           & 79.4          & 34.2 & 61.81          & 68.7           & 86.3           & 74.1           &31.3 & 63.61           & 72.1           & 83.6           & 72.3         &27.9  & 59.71            \\ \hline  \hline
\multirow{7}{*}{{\bf Reduce}} & {\bf Algorithm}      & {\bf Acc.}     & {\bf Pre.}     & {\bf Rec.}     & {\bf Err.}  & {\bf Time}     & {\bf Acc.}     & {\bf Pre.}     & {\bf Rec.}     & {\bf Err.} & {\bf Time}      & {\bf Acc.}     & {\bf Pre.}     & {\bf Rec.}     & {\bf Err.}  & {\bf Time}     \\ \cline{2-17}
                                   &
{\bf Tree}           & 72.4           & 95.3           & 69.3           & 27.6 & 13.51           & 72.8           & 92.2           & 73.0           & 27.8& 10.23            & 63.2           & 83.4           & 68.7           &36.8 & 13.25           \\ \cline{2-17}
                                   &
{\bf Boost}          & 83.5           & 93.4           & 85.1          & 16.5 & 297.29          & 93.1           & 98.7           & 92.3         &6.9  & 269.37          & 66.8           & 91.7           & 54.7           &33.2 & 198.37         \\ \cline{2-17}
                                   &
{\bf Glm}            & 61.9           & 92.6           & 65.1          & 38.1 & 17.32         & 77.2           & 91.3           & 75.3           &22.8 & 17.39           & 68.7           & 91.9           & 71.9       &  31.3  & 19.83           \\ \cline{2-17}
                                   &
{\bf CTree}          & 79.3           & 92.6           & 81.4          & 20.7& 16.85     &       81.4           & 91.1           & 81.3 & 18.6& 16.52      &      61.7           & 91.5           & 65.1          &38.3  & 17.13          \\ \cline{2-17}
                                   &
{\cellcolor[gray]{.8} \bf R.F.}  & {\cellcolor[gray]{.8} \bf  95.3} & {\cellcolor[gray]{.8} \bf  98.1} & {\cellcolor[gray]{.8} \bf  95.9} & {\cellcolor[gray]{.8} \bf 4.7 } & {\cellcolor[gray]{.8} \bf  35.65}  & {\cellcolor[gray]{.8} \bf  92.5} & {\cellcolor[gray]{.8} \bf  97.6} & {\cellcolor[gray]{.8} \bf  92.4} & {\cellcolor[gray]{.8} \bf 7.5 } & {\cellcolor[gray]{.8} \bf  28.54}  & {\cellcolor[gray]{.8} \bf 83.4} & {\cellcolor[gray]{.8} \bf  91.5} & {\cellcolor[gray]{.8} \bf  95.6} & {\cellcolor[gray]{.8} \bf 16.6 } & {\cellcolor[gray]{.8} \bf  27.93}  \\ \cline{2-17}
                                   &
{\bf N.N.} & 74.5           & 91.5           & 75.3         &   25.5   & 89.74       & 81.5           & 97.3           & 71.6 & 18.5 & 78.61           & 74.9           & 94.1          & 83.7         & 25.1  & 85.37          \\ \hline
\end{tabular}
\vspace{-5pt}
\end{table*}

\subsubsection{Performance Evaluation of the Schedulers}
\begin{figure*}[ht]
\minipage{0.32\textwidth}
\scriptsize
  \includegraphics[scale=0.55]{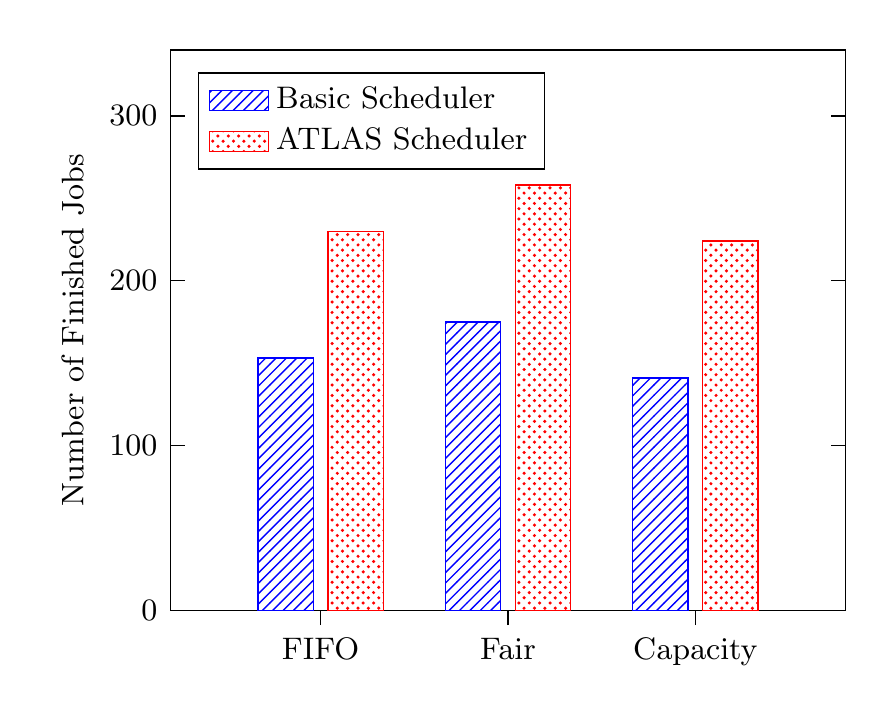}
  \caption{\scriptsize{Finished Jobs}}\label{fig:finsihedjobs}
\endminipage\hfill
\minipage{0.32\textwidth}
  \includegraphics[scale=0.55]{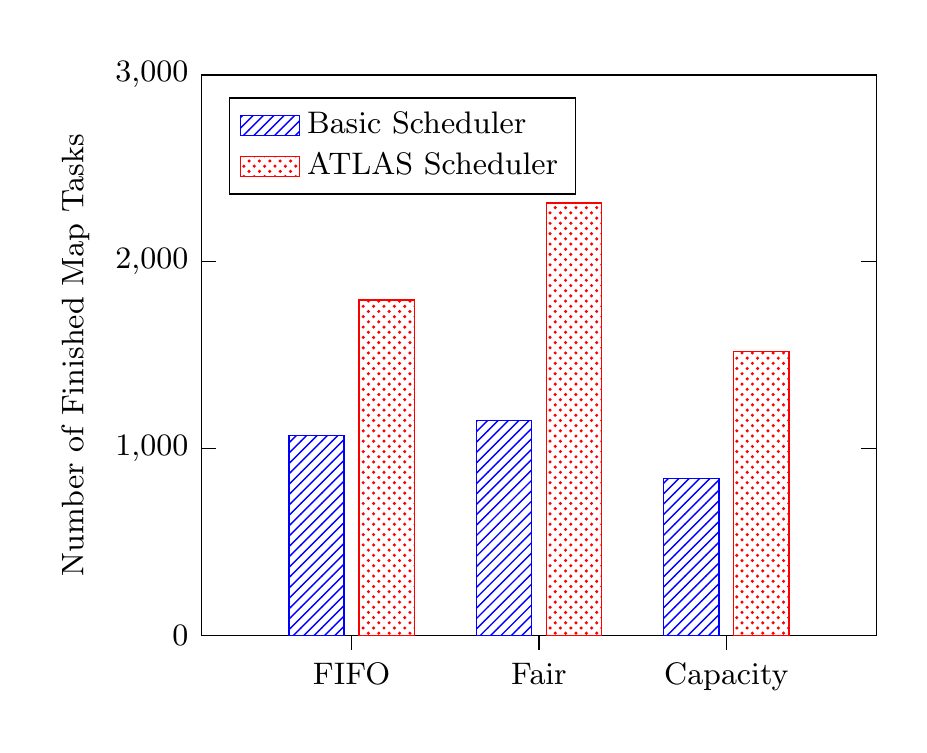}
  \caption{\scriptsize{Finished Map Tasks}}\label{fig:finsihedmaptasks}
\endminipage\hfill
\minipage{0.32\textwidth}%
  \includegraphics[scale=0.55]{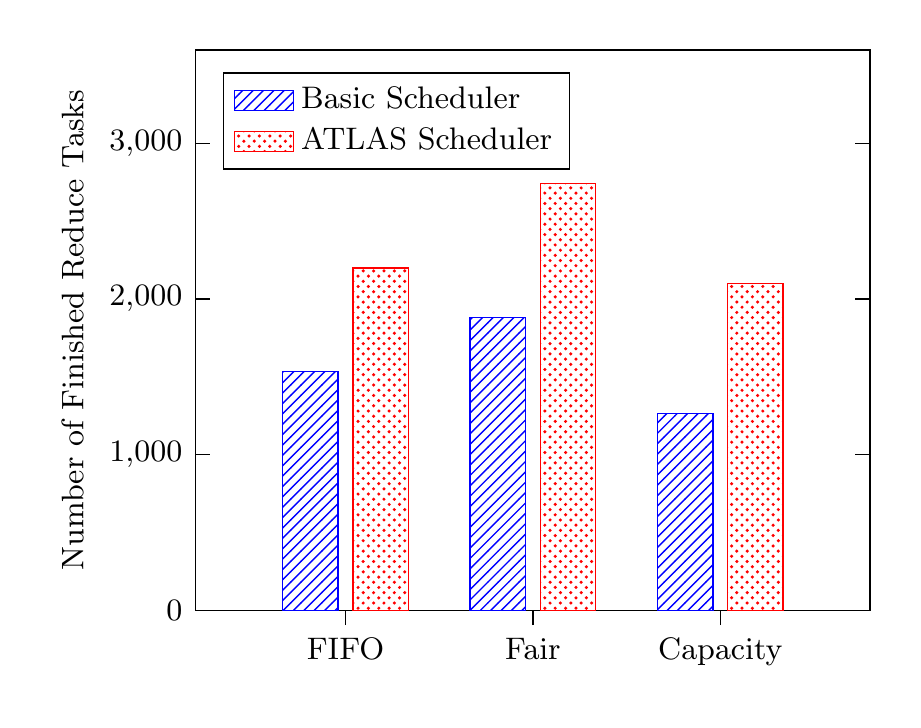}
  \caption{\scriptsize{Finished Reduce Tasks}}\label{fig:finishedreducetasks}
\endminipage
\vspace{-5pt}
\minipage{0.32\textwidth}
  \includegraphics[scale=0.55]{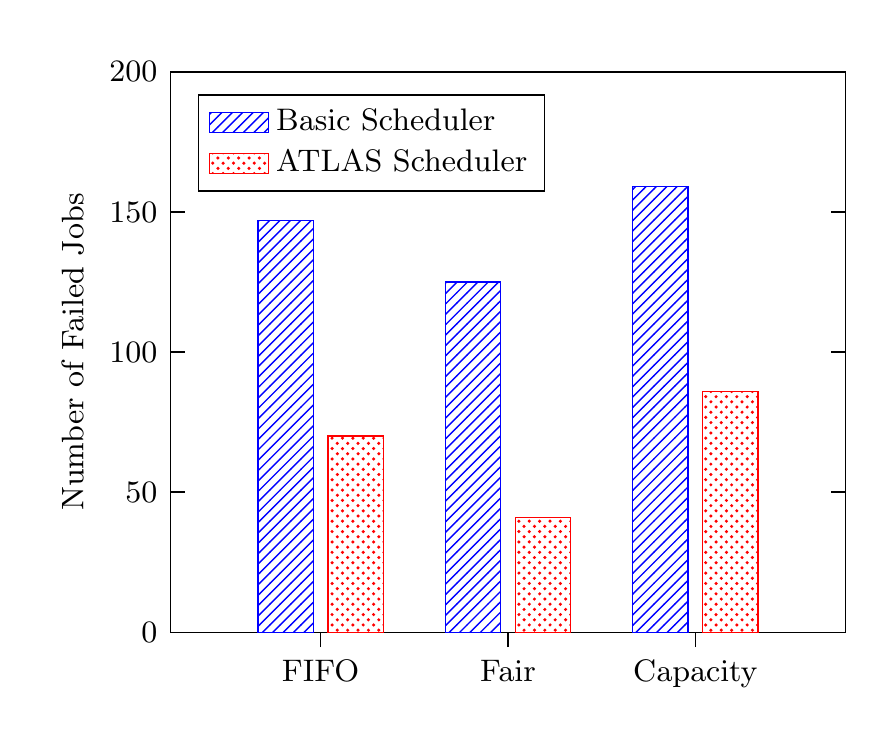}
  \caption{\scriptsize{Failed Jobs}}\label{fig:failedjobs}
\endminipage\hfill
\minipage{0.32\textwidth}
  \includegraphics[scale=0.55]{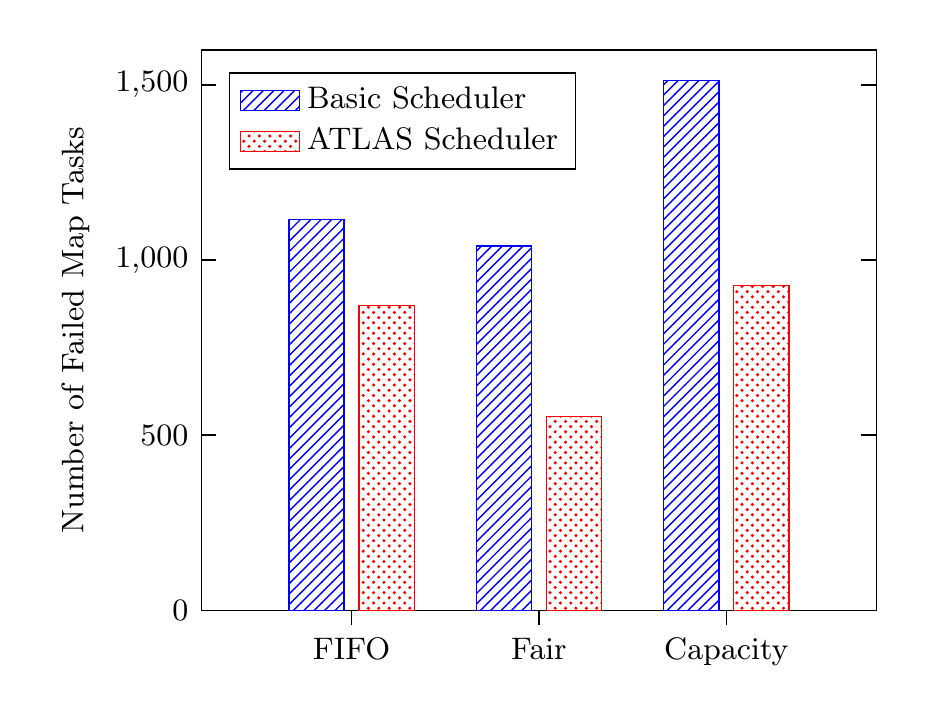}
  \caption{\scriptsize{Failed Map Tasks}}\label{fig:failedmaptasks}
\endminipage\hfill
\minipage{0.32\textwidth}%
  \includegraphics[scale=0.55]{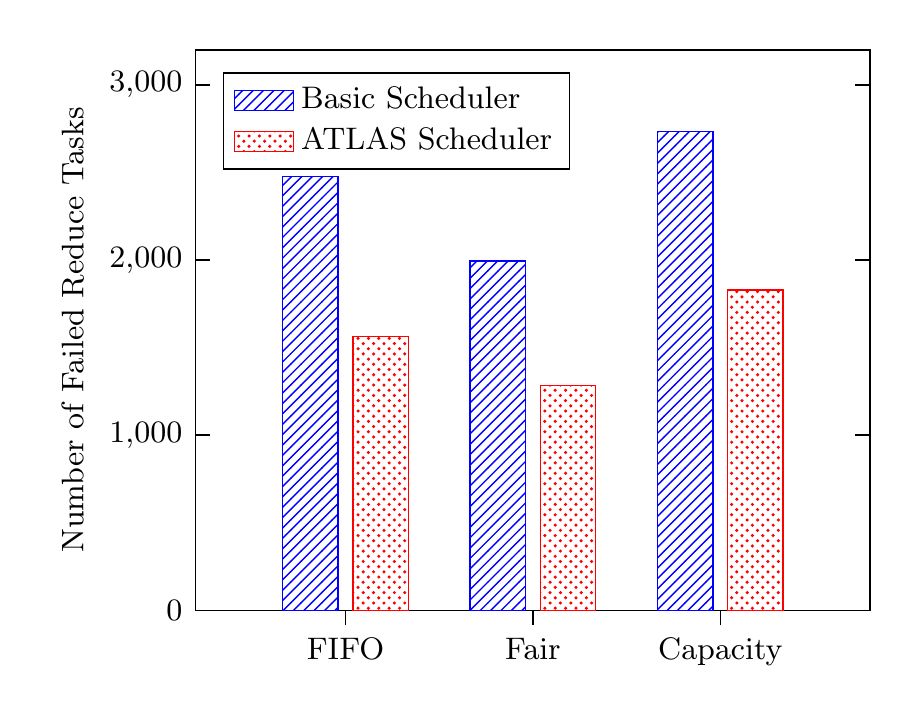}
  \caption{\scriptsize{Failed Reduce Tasks}}\label{fig:failedreducetasks}
\endminipage
\vspace{-5pt}
\minipage{0.32\textwidth}
  \includegraphics[width=50mm,height=45mm]{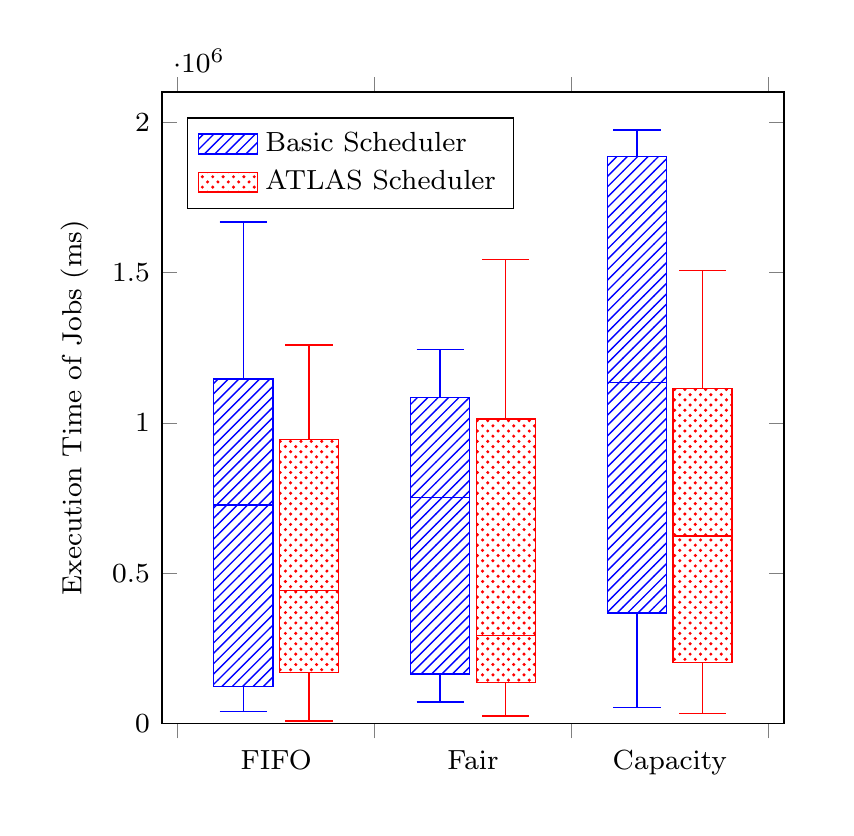}
  \caption{\scriptsize{Jobs Exec. Time}}\label{fig:execjobs}
\endminipage\hfill
\minipage{0.32\textwidth}
  \includegraphics[width=50mm,height=45mm]{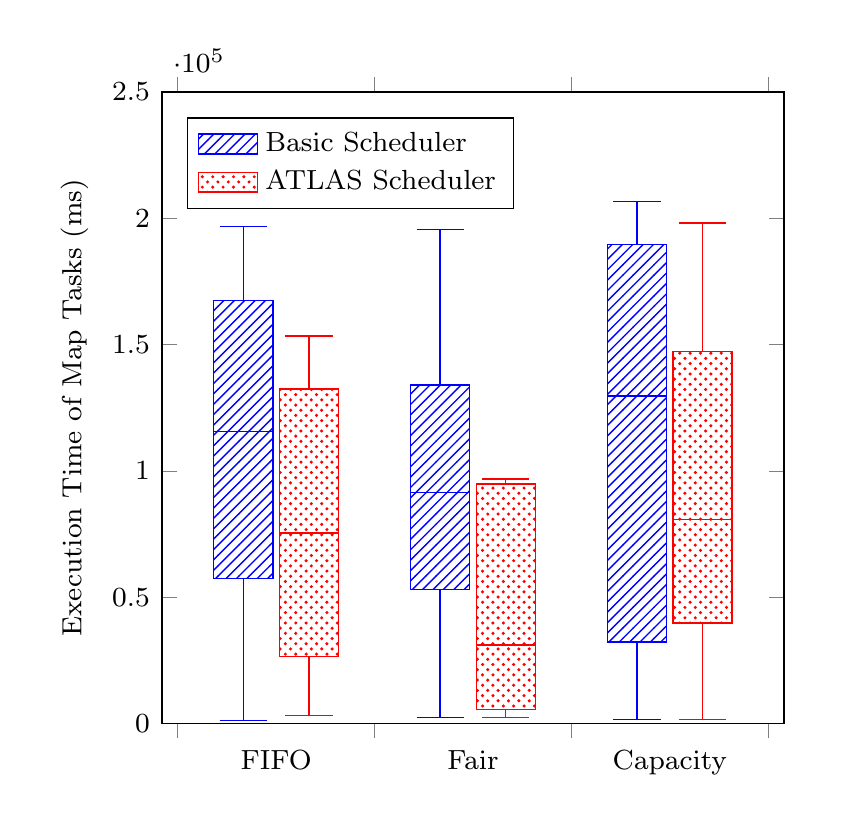}
  \caption{\scriptsize{Map Exec. Time}}\label{fig:execmap}
\endminipage\hfill
\minipage{0.32\textwidth}%
  \includegraphics[width=50mm,height=45mm]{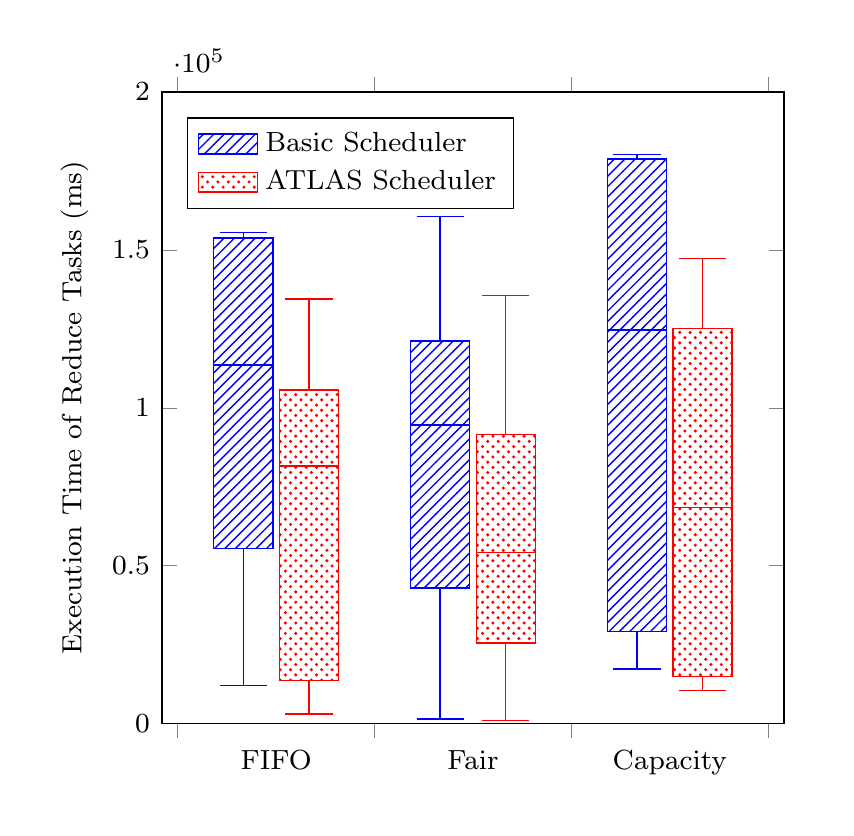}
  \caption{\scriptsize{Reduce Exec. Time}}\label{fig:execreduce}
\endminipage
\end{figure*}

As stated in Section~\ref{predictionresults}, we used the Random Forest algorithm (to predict the scheduling outcome of tasks) when implementing the proposed ATLAS scheduler. 
Figure~\ref{fig:finsihedjobs}, Figure~\ref{fig:finsihedmaptasks} and Figure~\ref{fig:finishedreducetasks} present respectively the number of finished jobs, map and reduce tasks for the three schedulers. Overall, we observe that the number of finished jobs, map and reduce tasks in ATLAS are higher in comparison to the results obtained for the FIFO, Fair, and Capacity schedulers. 
This was expected since the prediction model enables the quick rescheduling of tasks that are predicted to fail.
In addition, the improvement is larger for FIFO and Fair schedulers compared to the capacity scheduler. This happens because the capacity scheduler forces the killing of any tasks consuming more memory than configured~\cite{CapacityScheduler}. 
The number of finished tasks is improved by up to 46\% when using ATLAS instead of the Fair scheduler (see \textit{ATLAS-Fair} in Figure~\ref{fig:finsihedmaptasks}), and the number of finished jobs increased by 27\% when using ATLAS instead of the Fair scheduler (see \textit{ATLAS-Fair} in Figure~\ref{fig:finsihedjobs}).
The improvement of the number of finished jobs is lower than the improvement of the number of finished tasks since a single task failure causes the whole job to fail. 
We also noticed that the number of failed jobs and tasks was decreased by up to 28\% for the jobs (see \textit{ATLAS-Fair} in Figure~\ref{fig:failedjobs}) and up to 39\% for the tasks (see \textit{ATLAS-Capacity} in Figure~\ref{fig:failedmaptasks}) respectively. Moreover, we also observed that when the failure of one map task causes the failure of the dependent reduce tasks belonging to the same job, ATLAS is unable to propose a better scheduling decision (because some data are lost, as explained in Figure~\ref{Figure:exampleofjobfailure}).\\

\noindent Moreover, we noticed that the number of finished single and chained jobs was improved.
This was expected because ATLAS enables the successful processing of the tasks composing these jobs and because of the  dependency between the jobs within the chained jobs.
In addition, the number of finished single jobs was higher than the number of finished chained jobs. This is due to the dependency between the jobs composing the chained one (\ie{} sequential ones). Also, a single job failure in the composed chain can cause the failure of the whole chained job.

\noindent \textbf{In general, we conclude that our proposed ATLAS scheduler can reduce up to 39\% of tasks failures and up to 28\% of jobs failures that are experienced by the 3 other schedulers (\ie{} FIFO, Fair, and Capacity).}
\\

\noindent Overall, the execution time of tasks and jobs is lower for ATLAS. We attribute this outcome to the fact that \textit{ATLAS} reduces the number of launched attempts and the time spent to execute the attempts.
The total execution time of jobs was decreased on average by 10 minutes (from 20 minute to 10 minute), representing a 30\% of reduction of the total execution time of these jobs (see the \textit{ATLAS-Capacity} in Figure~\ref{fig:execjobs}) and the total execution time of tasks by about 1.33 minute (from 2.33 minute to 1 minute) (see \textit{ATLAS-Capacity}: reduce task in Figure~\ref{fig:execreduce}).
For long running jobs (running for 40-50 minutes), the reduction was up to 25 minutes (representing a 54\% reduction) over the Capacity scheduler.
In this context, we should mention that there was an overhead associated with the training phase of the predictive algorithm and the communication between the JobTracker and TaskTrackers. However, this overhead was very small and is included in the execution time of ATLAS presented on Figure~\ref{fig:execjobs}, Figure~\ref{fig:execmap}, and Figure~\ref{fig:execreduce}. In fact, the reduction in the number of failures largely compensated for this overhead.

\noindent \textbf{\textit{ALTLAS} successfully reduces the overall execution times of tasks and jobs in Hadoop.} 

\noindent By enabling the rescheduling of potential failed tasks in advance, ATLAS reduces the total execution time of the tasks and the number of tasks and jobs failure events. Consequently, it is expected that it will reduce resource utilisations in the cluster, since it should save the amount of resources that would have been consumed by tasks failed attempts.  
The results presented in Table~\ref{tab:resourcesutilisation} confirms this anticipated outcome.
In fact, by quickly rescheduling tasks predicted to be failed, ATLAS can save the resources that would have been consumed by these tasks. ATLAS speculatively executes the tasks predicted to be failed on multiple nodes to increase their chance of success. Whenever one of these tasks achieves a satisfactory progress, the other speculative executions are stopped.

\noindent \textbf{Overall, the jobs and tasks executed using ATLAS scheduling policies consumed less resources than those executed using the FIFO, Fair, or Capacity schedulers (in terms of CPU, Memory, HDFS Read and Write).}

\begin{table*}[ht!]
\centering \scriptsize
\caption{Resources Utilisation of the Different Hadoop Schedulers}
\vspace{-5pt}
\label{tab:resourcesutilisation}
\begin{tabular}{|p{1.2cm}|p{4cm}|p{0.9cm}|p{0.9cm}|p{0.9cm}|p{0.9cm}|p{0.9cm}|p{0.9cm}|}
\hline
\multirow{3}{*}{{\bf Job/Task}} & \multirow{2}{*}{{\bf Scheduler}} & \multicolumn{2}{c|}{{\bf FIFO}} & \multicolumn{2}{c|}{{\bf Fair}} & \multicolumn{2}{c|}{{\bf Capacity}} \\ \cline{3-8}
                                &                                  & {\bf Basic}    & {\bf ATLAS}    & {\bf Basic}    & {\bf ATLAS}    & {\bf Basic}       & {\bf ATLAS}     \\ \cline{2-8}
                                & {\bf Resource}                   & {\bf Avg.}     & {\bf Avg.}     & {\bf Avg.}     & {\bf Avg.}     & {\bf Avg.}        & {\bf Avg.}      \\ \hline\hline
\multirow{4}{*}{{\bf Job}}      & {\bf CPU (ms)}                        & 11495           & 8415           & 12647           & 9538           & 14475             & 10784            \\ \cline{2-8}
                                & {\bf Memory ($10^{5}$ bytes}                     & 7479     & 4530     & 7741     & 3647  & 9463        & 5486      \\ \cline{2-8}
                                & {\bf HDFS Read ($10^{3}$ bytes)}                  & 9930     & 7431     & 10968     & 8762  & 12463       & 8360      \\ \cline{2-8}
                                & {\bf HDFS Write ($10^{3}$ bytes)}                 & 8583     & 5985     & 9784     & 6202     & 10285        & 7420     \\ \hline\hline
\multirow{4}{*}{{\bf Task}}     & {\bf CPU (ms)}                        & 3855           & 2520           & 4033           & 2184           & 4170              & 2851            \\ \cline{2-8}
                                & {\bf Memory ($10^{5}$ bytes}                     & 1412     & 1058      & 2496     & 1741     & 2638        & 2115     \\ \cline{2-8}
                                & {\bf HDFS Read ($10^{5}$ bytes)}                  & 1638     & 1215     & 1894     & 1428     & 7426        & 4541      \\ \cline{2-8}
                                & {\bf HDFS Write ($10^{5}$ bytes)}                 & 1774     & 1385     & 3643     & 2429  & 5052         & 3715      \\ \hline
\end{tabular}
\vspace{-5pt}
\end{table*}

%% file: threats.tex
\section{Threats To Validity}
\label{sec:threats}
\vspace{-2pt}
This section discusses the threats to validity of our study following the guidelines for case study research~\cite{robert2002case}.\\

\noindent \emph{\textbf{Construct validity threats}} concern the relation between theory and observation. 
When building the predictive models used by ATLAS, we did not include information about the requested resources by tasks, since this information was not available in the collected logs. However, it is possible that some tasks were failed because they did not receive their requested resources. Our predictive models would hardly predict such failures. Nevertheless, we used the number of available slots in the machines to predict task failure in case of shortage of resources. Hence, ATLAS can reschedule the task on a different machine (with enough resources). \\

\noindent \textbf{\emph{Internal validity threats}} concern the tools used to implement ATLAS.
In addition, we used AnarchyApe \cite{Faghri-Failure2012} to inject different types of failures in Hadoop machines. We relied on rate of failures observed in Google clusters \cite{Mbarka15}. 
It is possible that the majority of Hadoop clusters do not experience such high rate of failures. 
It is also very possible that our simulations missed some types of failures occurring in Hadoop clusters. Future works should be performed on a more diverse set of Hadoop clusters and different failure rates.\\ 

\noindent \textbf{\emph{Conclusion validity threats}} concern the relation between the treatment and the outcome. 
We implemented a procedure to check the time spent by ATLAS in a way to not exceed the time-out value specified by the scheduler. 
In addition, we carefully checked the time spent by the algorithm to check the availability of the TaskTarckers and the DataNodes and to activate them through the JobTracker in a way that do not generate extra overhead times. We also verified that the scheduling decisions generated by ATLAS did not violate any property of the system.\\

\noindent \textbf{\emph{Reliability validity threats}} concern the possibility to replicate our study. We believe that our proposed approach can be reused on other cloud platforms such as Microsoft Azure~\cite{Azure}. To do that, a developer only needs to record the log files of processing nodes, build the predictive models, and implement the ATLAS algorithm on top of a Hadoop scheduler (like FIFO, Fair or Capacity) to adjust scheduling decisions according to task failure predictions.\\

\noindent \textbf{\emph{External validity threats}} concern the generalization of our obtained results. 
Further validation on larger clusters using diverse sets of tasks and jobs is desirable. 

%% file: relatedwork.tex
\section{Related Work}
\label{sec:relatedwork}
\subsection{Fault-Tolerance Mechanisms in Hadoop}
Hadoop was designed with some built-in fault-tolerance mechanisms. For example, the HDFS keeps multiple replicas of each data block on several different nodes, to allow for a quick restoration of data in the event of a node failure. Map and Reduce tasks are re-scheduled whenever they fail. Whenever a node fails, all the Map and
Reduce tasks on that node are re-scheduled to another node for re-execution. However, these simple redo solutions are not always effective. For example, If a failed task was almost completed, repeating the whole task can significantly increase the computation cost of the jobs. To address these limitations, researchers have proposed new mechanisms to improve the fault-tolerance of Hadoop. Quiane-Ruiz \emph{et al.} \cite{RAFTing-Quiane2011} proposed RAFT, a Recovery Algorithm for Fast-Tracking in Mapreduce that saves the states of tasks at multiple checkpoints. Whenever a task fails, the JobTracker can restart the task from the last checkpoint. They considered three types of checkpoints: (1) local checkpoints (RAFT-LC) to store states of the local reducer and mapper, (2) remote checkpoints (RAFT-RC) to store the intermediate results of remote workers, and (3) query metadata checkpoints (RAFT-QMC) to store the offset of input key-value pairs producing the intermediate results. RAFT does not re-execute the finished tasks belonging to a failed jobs. They assessed the performance of RAFT and found that it could reduce the total execution time of tasks by 23\% under different type of failures.
Qi \emph{et al.} \cite{Improving-Qi2014}, developed an algorithm called Maximum Cost Performance (MCP), to improve existing speculative execution strategies. 
However, this algorithm (\ie{} MCP) was found to negatively impacting the scheduling time of some jobs (batch jobs in particular)~\cite{DynamicMR2014}.
The CREST (Combination Re-Execution Scheduling Technology) algorithm was proposed to improve MCP, by using data locality during the speculative scheduling of slow running tasks~\cite{CREST-Lei2011}

\subsection{Adaptive Scheduling in Hadoop}
Given the dynamic nature of Hadoop environment, its scheduler can use information about the factors affecting their behavior to make better scheduling decisions and improve cluster performance. One scheduler that was proposed to do that is LATE~\cite{Improving-Zaharia2008}. LATE collects data about running tasks and assigns weights to tasks based on their progress. Using historical information about the weights assigned to tasks in the past, LATE prioritizes new tasks waiting to be executed. LATE was able to improve the execution time of jobs by a factor of 2 in large Hadoop cluster.
Quan \emph{et al.} proposed SAMR (Self-Adaptive MapReduce scheduling), a scheduler that uses hardware system information over time to estimate the progress of tasks and adjust the weights of map and reduce tasks, to minimize the total completion time of a job\cite{SAMR-Quan2010}. SAMR does not consider job characteristics such as size, execution time, or weights. To improve on this limitation of SAMR, Xiaoyu \emph{et al.} proposed ESAMR(Enhanced Self-Adaptive MapReduce scheduling)~\cite{ESAMR-Xiaoyu2012} which considers system information about straggling tasks, jobs length, etc. ESAMR uses the K-means clustering algorithm to estimate task execution times. 
In \cite{Aselfadaptive-Tang2015}, Tang \emph{et al.} proposed a scheduling algorithm named SARS (Self-Adaptive Reduce Start time) which uses job completion time, reduce completion time and the total completion time information as well as system information to decide on the starting time of reduce tasks. By improving the decisions about when to start reduce tasks, SARS could reduce the average response time of the tasks by 11\%.
Kc and Anyanwu~\cite{SchedulingKC2010} also proposed an algorithm that schedules jobs on Hadoop using user's specified deadlines. 
Li \emph{et al.}~\cite{WOHA-Li2014} proposed WOHA (WOrkflow over HAdoop) with the aim to improve workflow deadline satisfaction rates in Hadoop clusters. WOHO relies on job ordering and progress requirements to select the worklflow that falls furthest from its progress based on the Longest Path First and Highest Level First algorithms. Results show that WOHO can improve workflow deadline satisfaction rates in Hadoop clusters by 10\% compared to the existing scheduling solutions (FIFO, Fair and Capacity). 


%% file: conclusion.tex
\section{Conclusion and Future Work}
\label{sec:conclusion}
In this paper, we proposed ATLAS (AdapTive faiLure-Aware Scheduler), a new scheduler for Hadoop. The primary goal of ATLAS is to reduce the failure rates of jobs and tasks and their running times in Hadoop clusters. Based on information about events occurring in the cloud environment and statistical models, ATLAS can adjust its scheduling decisions accordingly, in order to avoid failure occurrences. We implemented ATLAS in Hadoop deployed on Amazon Elastic MapReduce (EMR) and performed a case study to compare its performance with those of the FIFO, Fair and Capacity schedulers.
Results show that ATLAS can reduce the percentage of failed jobs by up to 28\% and the percentage of failed tasks by up to 39\%. Although ATLAS requires training a predictive model, we found that the reduction in the number of failures largely compensates for the model training time. ATLAS could reduce the total execution time of jobs by 10 minutes on average, and by up to 25 minutes for long running jobs. ATLAS also reduces CPU and memory usages, as well as the number of HDFS Reads and writes.
In the future, we plan to extend ATLAS using unsupervised learning algorithms and assess the performance of ATLAS when the prediction model is retrained at fixed time intervals. 

%% file: IPCCC-TR.bbl
\begin{thebibliography}{10}
\providecommand{\url}[1]{#1}
\csname url@samestyle\endcsname
\providecommand{\newblock}{\relax}
\providecommand{\bibinfo}[2]{#2}
\providecommand{\BIBentrySTDinterwordspacing}{\spaceskip=0pt\relax}
\providecommand{\BIBentryALTinterwordstretchfactor}{4}
\providecommand{\BIBentryALTinterwordspacing}{\spaceskip=\fontdimen2\font plus
\BIBentryALTinterwordstretchfactor\fontdimen3\font minus
  \fontdimen4\font\relax}
\providecommand{\BIBforeignlanguage}[2]{{%
\expandafter\ifx\csname l@#1\endcsname\relax
\typeout{** WARNING: IEEEtran.bst: No hyphenation pattern has been}%
\typeout{** loaded for the language `#1'. Using the pattern for}%
\typeout{** the default language instead.}%
\else
\language=\csname l@#1\endcsname
\fi
#2}}
\providecommand{\BIBdecl}{\relax}
\BIBdecl

\bibitem{MapReduce-Dean2008}
J.~Dean and S.~Ghemawat, ``{MapReduce: Simplified Data Processing on Large
  Clusters},'' in \emph{ACM Communications}, 51(1):107--113, 2008.

\bibitem{AHybrid-Rasooli2012}
A.~Rasooli and D.~G. Down, ``{A Hybrid Scheduling Approach for Scalable
  Heterogeneous Hadoop Systems},'' in \emph{International Conference on SC
  Companion: High Performance Computing, Networking Storage and Analysis}, pp.
  1284--1291, 2012.

\bibitem{Dean-Experiences2006}
J.~Dean, ``{Experiences with MapReduce, an Abstraction for Large-scale
  Computation},'' in \emph{International Conference on Parallel Architectures
  and Compilation Techniques}, pp. 1--1, 2006.

\bibitem{Dinu-Understanding2012}
F.~Dinu and T.~E. Ng, ``{Understanding the Effects and Implications of Compute
  Node Related Failures in Hadoop},'' in \emph{International Symposium on
  High-Performance Parallel and Distributed Computing}, pp. 187--198, 2012.

\bibitem{Kim15}
Y.-P. Kim, C.-H. Hong, and C.~Yoo, ``{Performance Impact of JobTracker Failure
  in Hadoop},'' \emph{International Journal of Communication Systems},
  28(7):1265--1281, 2015.

\bibitem{Dinu-Hadoop2011}
F.~Dinu and T.~Ng, ``{Hadoop Overload Tolerant Design Exacerbates Failure
  Detection and Recovery},'' in \emph{International Workshop of Networking
  Meets Databases}, pp. 1--7, 2011.

\bibitem{Mbarka15}
M.~Soualhia, F.~Khomh, and S.~Tahar, ``{Predicting Scheduling Failures in the
  Cloud: A Case Study with Google Clusters and Hadoop on Amazon EMR},'' in
  \emph{International Conference on High Performance Computing and
  Communications}, pp. 58--65, 2015.

\bibitem{Hadoop-Jobs}
\BIBentryALTinterwordspacing
{Hadoop Jobs: WordCount, TeraSort and TeraGen}. [Online]. Available:
  \url{http://www.michael-noll.com/blog/2011/04/09/benchmarking-and-stress-testing-an-hadoop-cluster-with-terasort-testdfsio-nnbench-mrbench/,
  Last Access September, 2015}
\BIBentrySTDinterwordspacing

\bibitem{Par-Lee2012}
K.-H. Lee, Y.-J. Lee, H.~Choi, Y.~D. Chung, and B.~Moon, ``{Parallel Data
  Processing with MapReduce: A Survey},'' 40(4):11-20, 2012.

\bibitem{Abstract-Schultz2012}
J.~Schultz, J.~Vierya, and E.~Lu, ``Abstract: Analyzing patterns in large-scale
  graphs using mapreduce in hadoop,'' in \emph{SC Companion: High Performance
  Computing, Networking, Storage and Analysis (SCC)}, pp. 1457--1458, 2014.

\bibitem{Matchmaking-Cloud2011}
C.~He, Y.~Lu, and D.~Swanson, ``{Matchmaking: A New MapReduce Scheduling
  Technique},'' in \emph{International Conference on Cloud Computing Technology
  and Science}, pp. 40--47, 2011.

\bibitem{Dominant-Ghodsi2011}
A.~Ghodsi, M.~Zaharia, B.~Hindman, A.~Konwinski, S.~Shenker, and I.~Stoica,
  ``Dominant resource fairness: Fair allocation of multiple resource types,''
  in \emph{Proceedings of the 8th USENIX Conference on Networked Systems Design
  and Implementation}, ser. NSDI'11, pp. 24--24, 2011.

\bibitem{Improving-Yuting2013}
Y.~Ji, L.~Tong, T.~He, J.~Tan, K.~won Lee, and L.~Zhang, ``{Improving Multi-job
  MapReduce Scheduling in an Opportunistic Environment},'' in
  \emph{International Conference on Cloud Computing}, pp. 9--16, 2013.

\bibitem{Enhancement-Raj2012}
A.~Raj, K.~Kaur, U.~Dutta, V.~Sandeep, and S.~Rao, ``{Enhancement of Hadoop
  Clusters with Virtualization Using the Capacity Scheduler},'' in
  \emph{International Conference on Services in Emerging Markets}, pp. 50--57,
  2012.

\bibitem{KoMaking2010}
S.~Y. Ko, I.~Hoque, B.~Cho, and I.~Gupta, ``{Making Cloud Intermediate Data
  Fault-tolerant},'' in \emph{International Symposium on Cloud Computing}, pp.
  181--192, 2010.

\bibitem{[23]}
\BIBentryALTinterwordspacing
{The R Project for Statistical Computing}. [Online]. Available:
  \url{http://www.r-project.org/, Last Access September, 2015}
\BIBentrySTDinterwordspacing

\bibitem{AmazonInstances}
\BIBentryALTinterwordspacing
{Amazon EC2 Instances}. [Online]. Available:
  \url{http://aws.amazon.com/ec2/instance-types/, Last Access September, 2015}
\BIBentrySTDinterwordspacing

\bibitem{Faghri-Failure2012}
F.~Faghri, S.~Bazarbayev, M.~Overholt, R.~Farivar, R.~H. Campbell, and W.~H.
  Sanders, ``{Failure Scenario As a Service (FSaaS) for Hadoop Clusters},'' in
  \emph{International Workshop on Secure and Dependable Middleware for Cloud
  Monitoring and Management}, pp. 5:1--5:6, 2012.

\bibitem{CapacityScheduler}
\BIBentryALTinterwordspacing
Hadoop capacity scheduler. [Online]. Available:
  \url{http://hadoop.apache.org/docs/current1/capacity_scheduler.html, Last
  Access September, 2015}
\BIBentrySTDinterwordspacing

\bibitem{robert2002case}
R.~K. Yin, \emph{{Case Study Research: Design and Methods - Third Edition}},
  3rd~ed.\hskip 1em plus 0.5em minus 0.4em\relax SAGE Publication, 2002.

\bibitem{Azure}
\BIBentryALTinterwordspacing
Microsoft azure. [Online]. Available: \url{http://azure.microsoft.com/en-gb/,
  Last Access September, 2015}
\BIBentrySTDinterwordspacing

\bibitem{RAFTing-Quiane2011}
J.-A. Quiane-Ruiz, C.~Pinkel, J.~Schad, and J.~Dittrich, ``{RAFTing MapReduce:
  Fast recovery on the RAFT},'' in \emph{International Conference on Data
  Engineering}, pp. 589--600, 2011.

\bibitem{Improving-Qi2014}
Q.~Chen, C.~Liu, and Z.~Xiao, ``{Improving MapReduce Performance Using Smart
  Speculative Execution Strategy},'' \emph{IEEE Transactions on Computers},
  63(4):954-967, 2014.

\bibitem{DynamicMR2014}
S.~Tang, B.-S. Lee, and B.~He, ``{DynamicMR: A Dynamic Slot Allocation
  Optimization Framework for MapReduce Clusters},'' \emph{IEEE Transactions on
  Cloud Computing}, 2(3):333-347, 2014.

\bibitem{CREST-Lei2011}
L.~Lei, T.~Wo, and C.~Hu, ``{CREST: Towards Fast Speculation of Straggler Tasks
  in MapReduce},'' in \emph{International Conference on e-Business
  Engineering}, pp. 311--316, 2011.

\bibitem{Improving-Zaharia2008}
M.~Zaharia, A.~Konwinski, A.~D. Joseph, R.~Katz, and I.~Stoica, ``{Improving
  MapReduce Performance in Heterogeneous Environments},'' in
  \emph{International Conference on Operating Systems Design and
  Implementation}, pp. 29--42, 2008.

\bibitem{SAMR-Quan2010}
Q.~Chen, D.~Zhang, M.~Guo, Q.~Deng, and S.~Guo, ``{SAMR: A Self-adaptive
  MapReduce Scheduling Algorithm in Heterogeneous Environment},'' in
  \emph{International Conference on Computer and Information Technology}, pp.
  2736--2743, 2010.

\bibitem{ESAMR-Xiaoyu2012}
X.~Sun, C.~He, and Y.~Lu, ``{ESAMR: An Enhanced Self-Adaptive MapReduce
  Scheduling Algorithm},'' in \emph{International Conference on Parallel and
  Distributed Systems}, pp. 148--155, 2012.

\bibitem{Aselfadaptive-Tang2015}
Z.~Tang, L.~Jiang, J.~Zhou, K.~Li, and K.~Li, ``{A Self-Adaptive Scheduling
  Algorithm for Reduce Start Time},'' \emph{Future Generation Computer
  Systems}, 34-44(0):51-60, 2015.

\bibitem{SchedulingKC2010}
K.~Kc and K.~Anyanwu, ``{Scheduling Hadoop Jobs to Meet Deadlines},'' in
  \emph{International Conference on Cloud Computing Technology and Science},
  pp. 388--392, 2010.

\bibitem{WOHA-Li2014}
S.~Li, S.~Hu, S.~Wang, L.~Su, T.~Abdelzaher, I.~Gupta, and R.~Pace, ``{WOHA:
  Deadline-Aware Map-Reduce Workflow Scheduling Framework over Hadoop
  Clusters},'' in \emph{International Conference on Distributed Computing
  Systems}, pp. 93--103, 2014.

\end{thebibliography}
